\documentclass[aps,twocolumn,superscriptaddress,showpacs,showkeys,floatfix]{revtex4}
\usepackage{epsfig}
\usepackage{graphicx}
\usepackage{amsmath}
\usepackage[english]{babel}
\usepackage{multirow}
\usepackage{placeins}

\begin{document}

\title{Characterizing the community structure of complex networks}

\author{Andrea Lancichinetti}
\affiliation{Complex
  Networks and Systems, ISI Foundation, Torino, Italy}
\author{Mikko Kivela}
\affiliation{Department of Biomedical Engineering and Computational
  Science, Aalto University, Espoo, Finland}
\author{Jari Saram\"aki}
\affiliation{Department of Biomedical Engineering and Computational
  Science, Aalto University, Espoo, Finland}
\author{Santo Fortunato}
\affiliation{Complex Networks and
  Systems, ISI Foundation, Torino, Italy}

\begin{abstract}

Community structure is one of the key properties of complex networks
and plays a crucial role in their topology and function. While an
impressive amount of work has been done on the issue of community detection, 
very little attention has been so far devoted to the investigation of communities in real
networks. We present a systematic empirical analysis of the
statistical properties of communities in large
information, communication, technological, biological, and social networks.
We find that the mesoscopic organization of networks of the same category
is remarkably similar. This is reflected in several characteristics
of community structure, which can be used as 
 ``fingerprints'' of specific network categories. 
 While community size distributions are always broad, 
certain categories of networks consist mainly of tree-like communities,
while others have denser modules. Average path lengths within communities
initially grow logarithmically with community size, but the growth saturates or 
slows down for communities larger than a characteristic size. This behaviour is related to the presence of hubs within communities,
whose roles differ across categories. Also the community embeddedness of 
nodes, measured in terms of the fraction of links within their communities,
has a characteristic distribution for each category. Our findings are verified by the use
of two fundamentally different community detection methods.

\end{abstract}

\maketitle

\section{Introduction}
The modern science of complex systems has experienced a significant
advance after the discovery that the graph representation of such
systems, despite its simplicity, reveals a set of crucial features that
suffice to disclose their general structural
properties, function and evolution
mechanisms~\cite{albert00,boccaletti06,newman03,dorogovtsev01,pastor04,barrat08,caldarelli07}.
Representing a complex system as a graph means turning the elementary
units of the system into nodes, while links between nodes indicate their mutual
interactions or relations. 
Many complex networks are characterized by a broad
distribution of the number of neighbors of a node, \emph{i.e.}~its degree. This is
responsible of peculiar properties such as high robustness against
random failures~\cite{cohen00} and the absence of a threshold for the spreading
of epidemics~\cite{pastor01}. 

Another important feature of complex networks is represented by their
mesoscopic structure, characterized by the presence of groups of
nodes, called communities or modules,  with a high
density of links between nodes of the same group and a comparatively
low density of links between nodes of different groups~\cite{girvan02,danon07,porter09,fortunato10}.
This compartmental organization of networks is very common in
systems of diverse origin. It was remarked already in the 1960's that a hierarchical
modular structure is necessary for the robustness and stability of complex
systems, and gives them an evolutionary advantage~\cite{simon62}.


Exploring network communities is important for three main reasons:
1) to reveal network organization at a coarse level, which may
help to formulate realistic mechanisms for its genesis and evolution;
2) to better understand dynamic processes taking place on the
network (e.g., spreading processes of epidemics and
innovation), which may be considerably affected by the modular
structure of the graph;  
3) to uncover relationships between the nodes which are not apparent
by inspecting the graph as a whole and which can typically be attributed
to the function of the system.

Therefore it is not surprising that the last years have witnessed an
explosion of research on community structure in graphs. The main
problem, of course, is how to detect communities in the first place,
and this is the essential issue tackled by most papers on the topic which have appeared in the literature.
A huge number of methods and techniques have been designed, but the
scientific community has not yet agreed on which methods are most
reliable and when a method should or should not be adopted. This is
due to the fact that the concept of community is ill-defined. 
Since the focus has been on method development,
very little has been done so far to address a fundamental
question of this endeavor: {\it what do
communities in real networks look like?} This is what we will try to
assess in this paper.

Previous investigations have shown that across a wide range of networks, the distribution of community
sizes is broad, with many small communities coexisting with some much
larger 
ones~\cite{palla05,newman04, danon07, clauset04, radicchi04}.
The tail of the distribution can be often quite well fitted by a power
law. 
Leskovec et
al.~\cite{leskovec08} have carried out a thorough investigation of the
quality of communities in real networks, measured by
the conductance score~\cite{bollobas98}. They 
found that the lowest conductance, indicating well-defined modules,  is attained
for communities of a characteristic size of $\sim100$ nodes, whereas
much larger communities are more ``mixed'' with the rest of the
network. For this reason they suggest that the mesoscopic 
organization of networks may have a core-periphery structure, where
the periphery consists of small well-defined communities and
the core comprises larger modules, which are more densely
connected to each other and therefore harder to detect. 
Guimer\'a and Amaral have proposed a classification of
the nodes based on their roles within communities~\cite{guimera05}. 

However, the fundamental properties of communities in real networks are
still mostly unknown. Uncovering such properties is
the main goal of this paper. For this purpose, we have performed an
extensive statistical analysis of the community structure of many real networks from nature, society and technology.
The main conclusion is that communities are characterized by
distinctive features, which are 
common for networks of the same class but differ from one class to
another. 
Remarkably, such characterization is independent of the specific
method adopted to find the communities.

\section{Data and methods}

As our target is to study the statistical features of communities, we need 
to employ data sets on large networks containing high numbers of 
communities of varying size. Our data sets contain
$\sim10^5-10^6$ nodes, with exception for protein interaction networks (PINs), where 
the largest available data sets are of the order of $10^4$ nodes.

\begin{table*}
\caption{\label{table1}  List of the network datasets used for
our analysis. For each network we specify the number of nodes and
links, the average and maximum degree.}
\begin{tabular}{ | l | l | l | l | l | l | l | }
\hline
\multicolumn{6}{|c|}{Network statistics} \\ \hline

\multirow{1}{*}{Category}
& name & \# nodes  &\#  links & average degree  & max degree  \\ \hline

\multirow{2}{*}{Communication}
& wikitalk & 2,394,385 & 4,659,560 & 3.89 & 100,029  \\
& email & 265,214 & 364,481 & 2.75 & 7,636 \\
 \hline

\multirow{2}{*}{Internet} 
&caida & 26,475 & 53,381 & 4.03  & 2,628  \\
&dimes & 26,211 & 76,261 & 5.82  & 3,988  \\ 
\hline

\multirow{4}{*}{Information} 
&Web google & 875,713 & 4,322,050 & 9.87   & 6,332   \\
&arxiv & 27,770 & 352,285 & 25.37  & 2,468  \\
&amazon & 410,236 & 2,439,440 & 11.89 & 2,760 \\
&Web BS & 685,230 & 6,649,470 &  19.41 & 84,230 \\
\hline

\multirow{3}{*}{Biological} 
&dmela & 7,498 & 22,678 & 6.05 & 178 \\
&yeast & 1,870 & 2,203 & 2.36  & 56 \\
&human & 4,998 & 21,747 & 8.70  & 282 \\
\hline

\multirow{4}{*}{Social} 
&live j & 4,846,609 & 42,851,211& 17.68   &  20,333  \\
&epinions & 75,879 & 405,740 & 10.69  &  3,044\\
&last fm & 2,647,364 & 11,245,707  & 8.49  & 13,431   \\ 
&slashdot & 773,60 & 469,180 & 12.13 &  2,539  \\
\hline
\end{tabular}
\end{table*}

Table~\ref{table1} lists the network datasets we have used, along
with some basic statistics.
Most of them have been downloaded from the Stanford Large Network
Dataset Collection ({\tt http://snap.stanford.edu/data/}). Some
networks are originally directed (e.g., the Web graph), but we have
treated them as undirected. Further details on all networks can be
found in the Appendix.

Overall, we have considered five categories of networks:
\begin{itemize}
\item{{\bf Communication networks.} This class comprises the email
    network of a large European research institution,
    and a set of relationships between Wikipedia users communicating via
    their discussion pages. Note that in both cases, communication is not necessarily
    personal but involves, e.g., mass emails, and thus these networks cannot be considered as social networks.}
\item{{\bf Internet.} Here we have two maps of the Internet at the
    Autonomous Systems (AS) level, produced by the two main
    projects exploring the topology of the Internet:
    CAIDA ({\tt http://www.caida.org/}) and DIMES ({\tt http://www.netdimes.org/}).}
\item{{\bf Information networks.} This class includes a citation network of online preprints in
   {\tt  www.arxiv.org}, a co-purchasing network of items sold by
   {\tt www.amazon.com} and two samples of the Web graph, one representing the
   domains {\tt berkeley.edu} and {\tt stanford.edu} (Web-BS), the other was
   released by Google (Web-G).}
\item{{\bf Biological networks.} This class contains the PINs of
    three organisms: fruit fly ({\it Drosophila melanogaster}), yeast ({\it Saccharomyces cerevisiae})
    and man (\it{Homo sapiens}).}
\item{{\bf Social networks.} Here we considered four datasets: 
    a network of friendship relationships between users of the
    on-line community {\it LiveJournal} ({\tt www.livejournal.com});
    the set of trust relationships between users of the consumer review site {\tt
      epinions.com}; the friendship network of users of {\tt
      slashdot.org}; the friedship network of users
    of {\tt www.last.fm}.} 
\end{itemize}

\begin{figure}[b]
\begin{center}
\includegraphics[width=1.1\columnwidth]{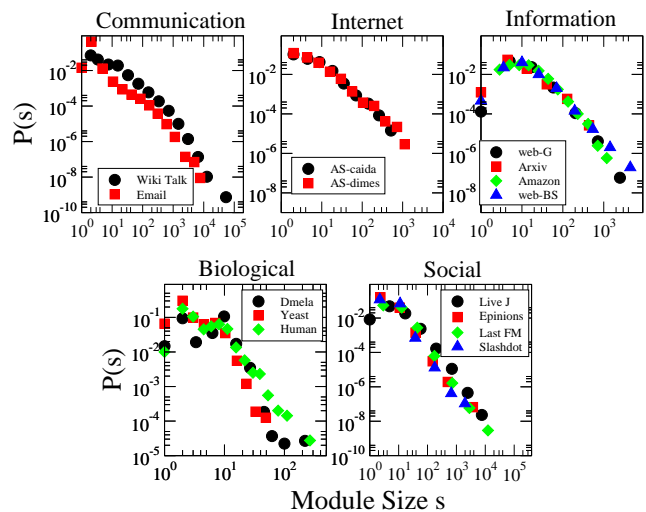}
\caption{\label{fig1} Distribution of community sizes. All distributions are
  broad, and similar for systems in the same category. Data points are averages within logarithmic bins
of the module size $s$.}
\end{center}
\end{figure}

The problem of choosing a method for detecting communities is a very delicate one. First, very efficient algorithms are needed, because the networks we study are large. This requirement rules out the majority of existing methods. Second, as discussed above, there is no common agreement on an all-purpose community detection method.
This is because of the absence of a shared definition of
community, which is justified by the nature of the problem itself. Consequently, there is also
 arbitrariness in defining
reliable testing procedures for the algorithms. 
Nevertheless, there
is a wide consensus on the definition of community originally
introduced in a paper by Condon and Karp~\cite{condon01}. The idea is
that a network has communities if the probability that two nodes of
the same community are connected exceeds the probability that
nodes of different communities are connected. This concept of
community has been implemented to create classes of benchmark graphs
with communities, such as those introduced by
Girvan and Newman~\cite{girvan02} and the graphs recently designed by  
Lancichinetti et al.~\cite{lancichinetti08}, which integrate the
benchmark by Girvan and Newman with realistic distributions of degree
and community size (LFR benchmark). 
Recent work indicates that some algorithms perform very well on the
LFR benchmark~\cite{lancichinetti09c}. In particular, the Infomap method
introduced by Rosvall and Bergstrom~\cite{rosvall08} has an outstanding
performance, and it is also fast and thus suitable for large networks. However, as every
community detection method has its own "flavor" and preference towards labeling certain
types of structure as communities, relying on a single method is not enough if general conclusions
on community structure are to be presented. Therefore we have cross-checked the results
obtained by Infomap with those produced by a very different algorithm, the Label
Propagation Method (LPM) proposed by Leung et al.~\cite{leung09}.  The
latter has proven to be reliable on the LFR benchmark and is also fast
enough to handle the largest systems of our collection. Detailed
descriptions of Infomap and the LPM are given in the Appendix. Here we just
point out the profound differences between the two techniques. Infomap
is a global optimization method, which aims to optimize a quality function
expressing the code length of an infinitely long random walk taking
place on the graph. The LPM is a local
method instead, where nodes are attributed to the same community where most of
their neighbors are. The partitions obtained by both methods
for the same network are in general different. However,
the general statistical features of community structure do not appear to depend
much on the details of partitions. In the following, only Infomap results will
be presented; for LPM, see Appendix.

\section{Results}

\begin{figure}[b]
\begin{center}
\includegraphics[width=1.1\columnwidth]{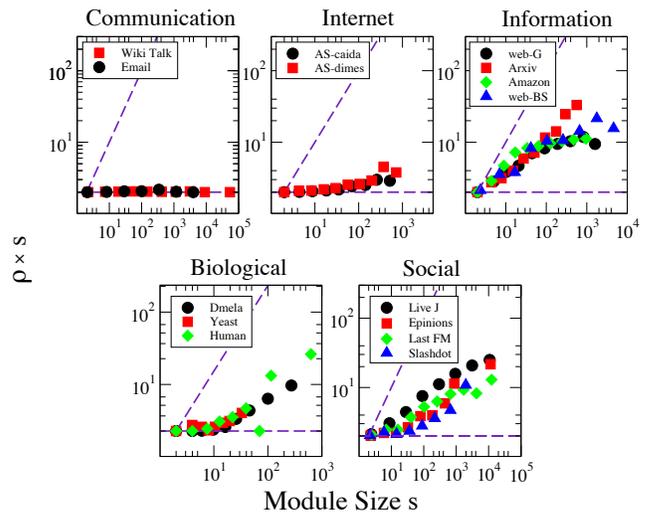}
\caption{\label{fig0} Scaled link density of communities as a function of the
community size. Communication and Internet networks consist of essentially
tree-like communities, while communities of social and information networks are
much denser. Small modules in biological networks are often tree-like, while larger modules
are denser. Data points are averages within logarithmic bins
of the module size $s$.}
\end{center}
\end{figure}
\begin{figure*}[t]
\begin{center}
\includegraphics[width=2.0\columnwidth]{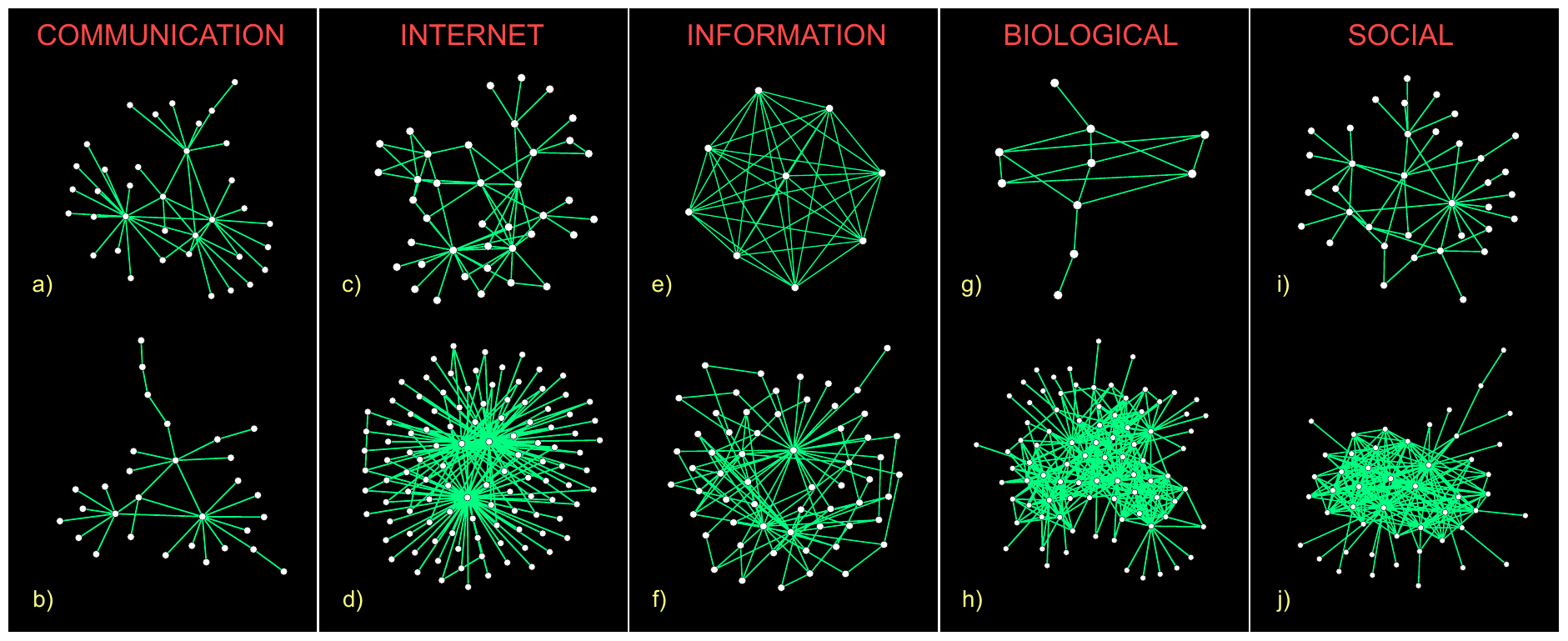}
\caption{Visualized examples of communities in networks of different classes. 
Communication networks (a: email, b: Wiki Talk) contain very sparse
communities with star-like hubs. These hubs give rise to very low shortest path lengths
within communities (see Fig.~\ref{fig2}). Star-like hubs are
are also present in Internet communities (c: DIMES, d: CAIDA), which are relatively
sparse as well. The CAIDA community displays a "merged-star" structure
fairly typical for these networks (see Appendix). 
On the contrary, information networks contain dense communities up to large
cliques (e: Amazon, f: Web-BS).  In biological networks, the larger the community, the
less tree-like it is (g: D. melanogaster, h: H. sapiens). Finally, communities in social
networks appear  on average fairly homogeneous (i: Slashdot, j: Epinions).}\label{commfig}
\end{center}
\end{figure*}

We begin the analysis by briefly discussing the distribution of community sizes (Fig.~\ref{fig1}).
We see that, as expected, for each system there is a wide range of
community sizes, spanning several orders of magnitude for the largest
systems. This is in agreement with earlier
studies~\cite{palla05,newman04, danon07, clauset04, radicchi04}. 
The overall shapes of the distributions are 
similar across systems of the same class. 
Distributions for biological networks show the largest differences,
which, however, is likely to result from noise as the networks are smaller. For biological
networks, analysis performed with the LPM shows slightly different, well overlapping distributions (see Appendix). 

Next, we turn to the topology of the communities, and
 study the link density of communities and its dependence on community size.
The link density of a subgraph is defined as the fraction of existing links to possible links, $\rho=2t/\left[s\left(s-1\right)\right]$,
where $t$ is the number of its internal links and $s$ its size measured in nodes.
Here, we use the scaled link density $\tilde{\rho}=\rho s = 2t/\left(s-1\right)$,
which also approximately amounts to the average community-internal
degree of nodes in the community. We have chosen this measure since
it clearly points out the nature of subgraphs. For trees, there are always
$s-1$ links, and hence $\tilde{\rho}_{tree}=2$. On the other hand, for full cliques
$\rho=1$ and hence $\tilde{\rho}_{clique}=s$.

Fig.~\ref{fig0} displays the average scaled link densities $\tilde{\rho}$
as function of community size for different networks. The dashed lines 
indicate the limiting cases ($\tilde{\rho}_{tree}=2, \tilde{\rho}_{clique}=s$).
We see that the link densities in the communication and Internet networks are very close to
the lower limit, which means that their communities are tree-like and contain only few or no loops. 
In communication networks, the scaled link density does not depend on
community size, whereas in Internet graphs large communities appear somewhat denser. 
Networks in these two classes are the sparsest in
our collection, as their very small average degree indicates that they are overall not
much denser than trees (see Table~\ref{table1}). It should be noted that in general, the 
intuitive view on communities is that they are "dense" compared to the rest of the network. However,
as the methods applied here yield partitions, the communities of a tree-like network are 
also necessarily tree-like.  Contrary to the above, the much denser information
networks reveal a different picture, where communities
are fairly dense objects, with the scaled density increasing
with $s$.  Especially in the Amazon network, communities with $s\lesssim10$ are almost cliques.
Social
networks show yet another pattern: the scaled density of the modules
grows quite regularly with the size $s$, approximately as a power law. 
Communities in social networks are mostly far from the two limiting cases: they are
denser than trees, but much sparser than cliques, with the exception
of small communities which appear more tree-like.
Finally, the biological networks are characterized
by two regimes: for $s\lesssim10$, communities are
very tree-like; for larger values of $s$ the scaled density increases with $s$. 
In Fig.~\ref{commfig} the characteristic communities of
each network class are illustrated.
\begin{figure}[b]
\begin{center}
\includegraphics[width=1.1\columnwidth]{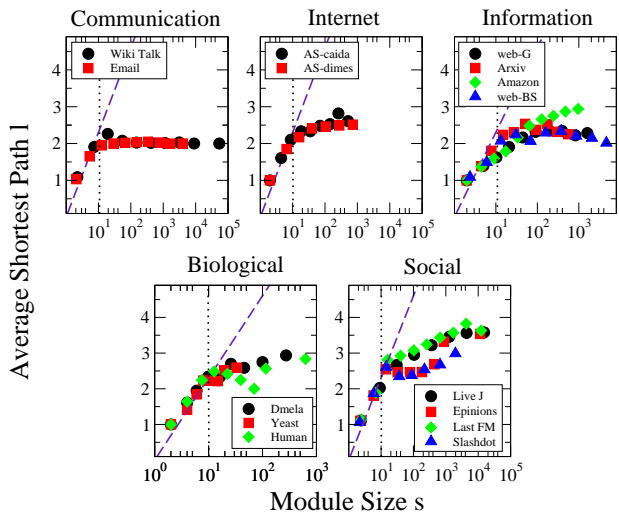}
\caption{\label{fig2} Average shortest path lengths $\ell$ within communities as a
  function of community size $s$. After an initial logarithmic ``small-world''
  regime (dashed diagonal line), the average shortest path grows much
  slower or saturates for communities with $s\gtrsim 10$ nodes (dotted vertical line). 
  Data points are averages within logarithmic bins
of module size $s$.}
\end{center}
\end{figure}

The compactness of communities can be measured using 
the average shortest path length $\ell$ within
each community. Fig.~\ref{fig2} displays the average values of $\ell$ as function of community size $s$. 
For all networks,  the average shortest path lengths $\ell$ are very small, $\ell<3$ with the exception
of social networks. 
Interestingly, all plots reveal the same basic
pattern, independently of the network class. 
For very small communities, $\ell$ grows
approximately as the logarithm of the community size (indicated by the
dashed line), which is the ``small-world'' property typically
observed in complex networks~\cite{watts98}. We call these 
modules {\it microcommunities}. For sizes $s$ of the
order of $10$, however, the increase of $\ell$
suddenly becomes less pronounced, and several curves reach a plateau.
Modules with $\gtrsim 10$ nodes are {\it macrocommunities}.  
The stabilization of the average shortest path length in macrocommunities can be 
attributed to the presence of nodes with high degree, \emph{i.e.}~hubs, which make
geodesic paths on average short. We remark that, since most of our systems have broad degree distributions,
shortest path lengths are very short~\cite{chung02}, but the sharp
transition we observe is unexpected and appears as an entirely novel feature.
 
For communication networks, there is a plateau with $\ell\sim 2$ for $s>10$.
As these communities are tree-like, this indicates that they have a star-like structure
where most nodes are connected to a central hub only and thus their distance equals two.
For the Internet networks, the joint presence of low density and low distances also means
that hubs dominate the structure -- here, "merged-star" structures consisting of two or more hubs
sharing many of their neighbors were observed (see
Fig.~\ref{commfig}d). This structure guarantees an efficient
communication between the systems' units. On the contrary,  information, social, and biological networks have a higher
density and hence their short path lengths are due to both the density and the presence of hubs.
Hubs play the least dominant role in social networks, as the average shortest path lengths keep
slowly increasing also for large $s$. 
\begin{figure}[b]
\begin{center}
\includegraphics[width=1.1\columnwidth]{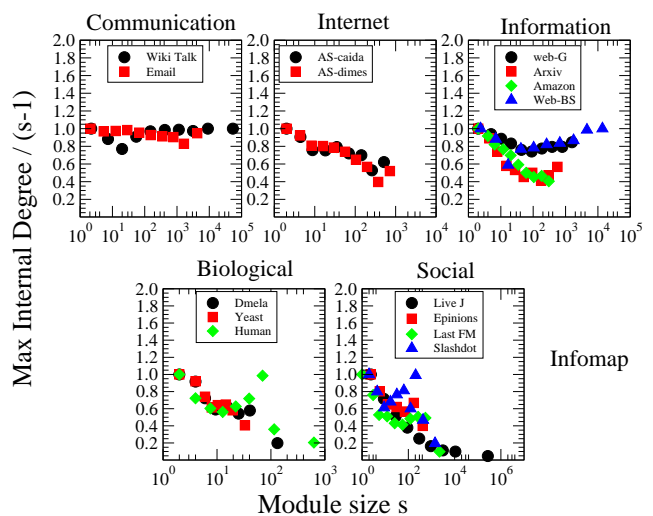}
\caption{\label{fig:maxk} The maximal observed internal degree of
  nodes as a function of the community size
$s$. This quantity equals one if any node is linked to all other nodes of its community, and thus quantifies
the dominance of hubs within communities.}
\end{center}
\end{figure}

The above picture is further corroborated by Fig.~\ref{fig:maxk},
which displays the ratio of the maximal observed 
community-internal degree of nodes $\max(k_{in})$ to $s-1$ as a
function of the community size $s$. This ratio equals unity if any node is connected to all other nodes in its community, and thus it quantifies the dominance of hubs within communities. For communication networks, $\max(k_{in})/(s-1)$ is close to unity even for large $s$, in accordance with the above observations on star-like communities. For Internet, this quantity somewhat decreases with $s$, as communities may contain multiple hubs which do not connect to all other nodes. In information networks, there are some differences. In the Web graphs, the largest communities contain nodes connecting (almost) the entire community. As the edge density in these communities is high, there may be several such nodes -- in a clique, all nodes have degree $s-1$. For biological
and social networks, there is a decreasing trend. Especially in social networks, there are few or no dominant hubs in large communities. 

\begin{figure}[t]
\begin{center}
\includegraphics[width=1.1\columnwidth]{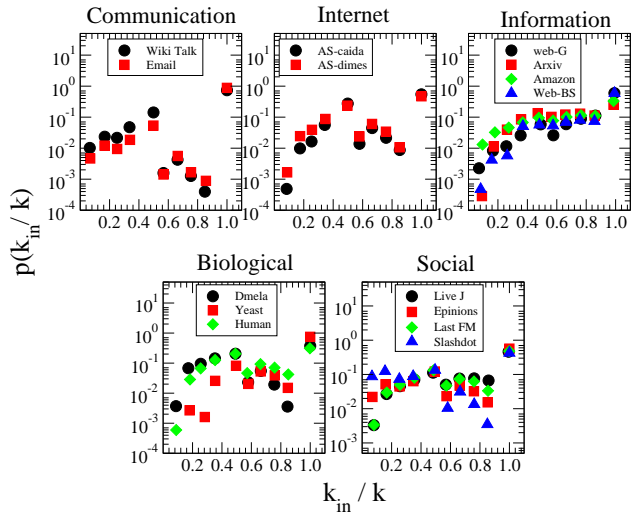}
\caption{\label{fig3} Probability distribution for $k_{in}/k$,
 the fraction of neighbors of a
  node belonging to its own community. Networks in the same
  class display similar behavior.  }
\end{center}
\end{figure}

Let us next take a closer look at the relationship between individual nodes and
community structure. Here, the most natural property to investigate
is the internal degree $k_{in}$, indicating the number of neighbors
of a node in its community. We measure the \emph{embeddedness} of a node in its
community with the ratio $k_{in}/k$, characterizing
the extent to which the node's neighborhood belongs to the same community as the node itself. 
The probability distribution of the embeddedness ratio of all nodes in their respective networks
is displayed in Fig.~\ref{fig3}.
One would straighforwardly assume that on average, the embeddedness of nodes would
be fairly large, and a substantial fraction of their neighbors should reside inside their
respective communities. However,  Fig.~\ref{fig3} shows a more intricate pattern,
where smaller values of $k_{in}/k$ are not at all rare. All of our networks
are characterized by a substantial fraction of nodes which are entirely internal
to their communities, \emph{i.e.} have no links to outside their community and thus $k_{in}/k=1$. 
These correspond to the rightmost data points in each plot, and such nodes typically
amount to over $50\%$ of all nodes. These nodes
have mostly a low degree (such as the degree-one nodes connected to hubs in communication networks).  Networks in the
same class follow essentially a very similar pattern. Communication networks
and the Internet have very similar-looking profiles, where the distribution has a peak
around  $k_{in}/k\sim 0.5$. 
 Information
networks, instead, have a rather different profile, with an initial
smooth increase reaching a plateau at about $k_{in}/k\sim 0.4$.
The biological networks, despite the inevitable noise, also show a
consistent picture across datasets. They somewhat resemble the communication and Internet networks,
with an initial rise until
$k_{in}/k\sim 0.5$, followed by a slow descent for larger
values. Social networks have a rather flat distribution over the whole
range, with little variations from one system to another. This means
that there are many nodes with most of their neighbors outside their own
community. Most community detection techniques, including the ones we have
adopted, tend to assign each node to the community which contains the
largest fraction of its neighbors. This implies that if a node has only a 
few neighbors within its own community, it will have even fewer neighbors
within other individual communities. Such nodes act as ``intermediates''
between many different modules, and are shared between
many communities rather than belonging to a single community only. Hence it would be
more correct to assign them to more than one community. Overlapping
communities are known to be very common in social networks, and 
dedicated techniques for their
detection have been introduced~\cite{palla05,zhang07,gregory07,lancichinetti09,evans09}.

%

%


\section{Discussion and conclusions}

Since the advent of the science of complex networks, its focus has shifted
from understanding the emergence and importance of system-level characteristics
to mesoscopic properties of networks. 
These are manifested in communities, \emph{i.e.} densely connected subgraphs.
Communities are ubiquitous in networks and typically play an important role in the 
function of a complex system -- modules in protein-interaction networks relate to specific
biological functions, and communities in social networks represent the fundamental
level of organization in a society. The dual problem of formally defining and accurately detecting 
communities has so far attracted the most of attention, at the cost of a lack of understanding
of the fundamental structural properties of communities. Our aim in this paper has been to uncover some
of these properties.

Our results indicate that communities detected in networks of the same class display surprisingly
similar structural characteristics. This is remarkable, as some classes are really broad and 
comprise systems of different origin (e.g. the class of information networks,
which includes graphs of citation, co-purchasing and the Web).
The result is verified by two different community detection methods which
are both partition-based but rely on entirely different principles. In accordance with earlier results, community size
distributions are broad for all systems we have studied. Link densities within communities depend
strongly on the network class. The average shortest path length displays similar behavior across
all classes, initially increasing logarithmically as a function of
community size (microcommunities) 
and then slowing down or saturating for communities of size
$s\gtrsim10$ (macrocommunities). 
In combination with our results on link density in communities,
the behavior of path lengths reveals a picture where high-degree nodes are very dominant in communities
of certain classes (communication, Internet) and play a less important
role in the connectivity of others, especially social networks. This
picture is corroborated by the analysis of maximal community-internal degrees of nodes.
Finally, also the probability distribution of the fraction of internal links for nodes displays a clear signature for each of the considered classes.

The signatures we have found are a sort of network
ID,  and could be used both to classify other systems and to identify new
network classes. Moreover, they could become essential elements of network
models, with the advantage of more accurate descriptions of real
networks and predictions of their evolution.

Although our results have been obtained using two different methods, their general validity merits
some discussion. As the concept of "community" is ill-defined, every method for detecting communities
is based on a specific interpretation of the concept.
Furthermore, the
underlying philosophies of methods can largely differ. 
Methods requiring that communities are "locally" very dense, such as clique percolation~\cite{palla05}, would detect 
only a few communities in the communication and Internet networks, as
they do not consider trees 
or stars as communities -- nevertheless, this result would be consistent for networks of the same class.
 On the other hand, it is evident that partition-based methods neglect
 the fact that nodes may participate in multiple communities. However,
 it is worth noting that whichever method is used, the resulting
 communities are actual subgraphs of the network under study,
 \emph{i.e.}~its building blocks. Thus their statistical properties
 reflect the mesoscopic organization of networks, and our results
 indicate that this organization is similar within classes of
 networks.

\begin{acknowledgments}

We gratefully acknowledge ICTeCollective. The project ICTeCollective acknowledges the financial support of the Future and Emerging
Technologies (FET) programme within the Seventh Framework Programme for Research of the
European Commission, under FET-Open grant number: 238597.

\end{acknowledgments}

\begin{appendix}

\section{Data sets: basic statistics}

\begin{figure}[h!]
\begin{center}
\includegraphics[width=\columnwidth]{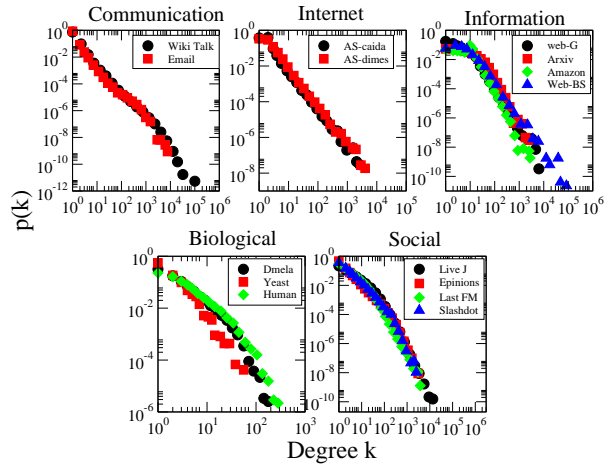}
\caption{Degree distributions.}
\label{degree_distr}
\end{center}
\end{figure}

\begin{figure}[h!]
\begin{center}
\includegraphics[width=\columnwidth]{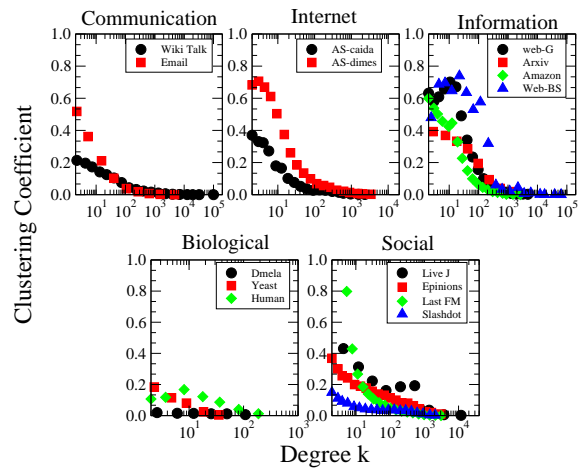}
\caption{Clustering coefficient versus degree.}
\label{clu_c}
\end{center}
\end{figure}

\begin{figure}[h!]
\begin{center}
\includegraphics[width=\columnwidth]{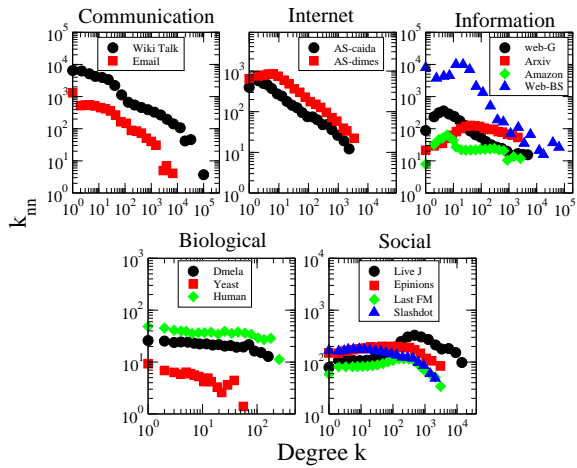}
\caption{Average nearest-neighbor degree $k_{nn}$ versus degree.}
\label{knn}
\end{center}
\end{figure}

In Fig.~\ref{degree_distr}  we show the degree
distributions for all the networks. The
degree distribution spans several orders of magnitude.
In Fig.~\ref{clu_c} the clustering coefficient~\cite{watts98} of nodes with degree
$k$ is plotted as a function of $k$, defined as the number of links between neighbors $t$ of the node divided
by the maximum possible number of such links f:
$c=t/(\frac{1}{2}k\left( k-1\right)) $.
As we can see, the shape of the clustering spectrum is
basically the same across all networks, with a rapid decrease of the clustering
coefficient with $k$, except for the Web graphs, which are known to
include very dense subgraphs and cliques, for which the clustering
coefficient can be appreciably high also for nodes of degree $\sim 100$. In Fig.~\ref{knn} we report
the average degree $k_{nn}$ of the neighbors of nodes with degree $k$
again as a function of $k$~\cite{vazquez01}. Communication networks, the Internet and
the Web graphs are clearly disassortative, the other networks are either
moderately disassortative or do not exhibit a particular
correlation. Only the Livejournal friendship network has an
assortative pattern for intermediate degree-values.

\section{The community detection methods}

In this section, we briefly explain the two community detection algorithms. For a detailed 
description, the reader is referred to the original publications.

Infomap~\cite{rosvall08} is based on the idea that a random walker exploring the
network should get trapped inside dense modules for a fairly long time, and cross
the boundaries of modules only infrequently. 
This simple idea is formalized by considering the problem of finding the
optimal description of the path of the walker, which can be achieved by labelling
every node with a prefix given by a unique name for the module it belongs to and a suffix given
by a unique name within its module. The labels of nodes, while unique within their
module, can be recycled in different modules to achieve the most
compressed description. According to such two-level description, given a
partition of the graph, one can compute the amount of information needed to describe the
path of the walker. If the network has a well-defined community
structure, the code length of the two-level description may be shorter
than the code length of the one-level description, in which each node
has a unique name, as the walker will
perform most of its steps within each module and comparatively few
between the modules. In this way, the recycling of the labels leads to
a more compact description of the process.
Then the problem of Infomap is finding the partition which gives the
smallest description length. This optimization problem is solved using a greedy optimization
algorithm in order to obtain the results in reasonable time. The use of random walks makes the
method naturally generalizable to the case of directed and weighted
graphs. For directed graphs, due to the possibility of having dangling
ends, which are sinks for the diffusion process, it is necessary to introduce a teleportation factor, similarly
to Google's PageRank algorithm~\cite{brin98}.

The Label Propagation Method~\cite{raghavan07} basically simulates the spreading of
labels based on the simple rule that at each iteration a given node takes the most
frequent label in its neighborhood. The starting configuration is chosen such that every
node is given a different label and the procedure is iterated until convergence. This
method has the problem of partitioning the network such that there are very big clusters, due to the
possibility of a
few labels to propagate over large portions of the graph. 
The LPM version that we used in our analysis is a modification by Leung et
al.~\cite{leung09} that handles this problem by introducing
a hop score which tells how far a certain label is from its origin. The hop
score is decreased while the label spreads through the network and
this improves the quality of the partitions found by the method.

\section{Main results from the Label Propagation Method}

In order to verify that our results are not due to the method alone, but 
represent real features of the mesoscopic organization of the networks,
we have carried the analysis presented for Infomap in the main paper with the Label
Propagation method as well. The following plots show the characteristics presented 
in the main paper obtained via the label propagation method. 
Results are consistent with those obtained with Infomap.

\begin{figure}[h!]
\begin{center}
\includegraphics[width=\columnwidth]{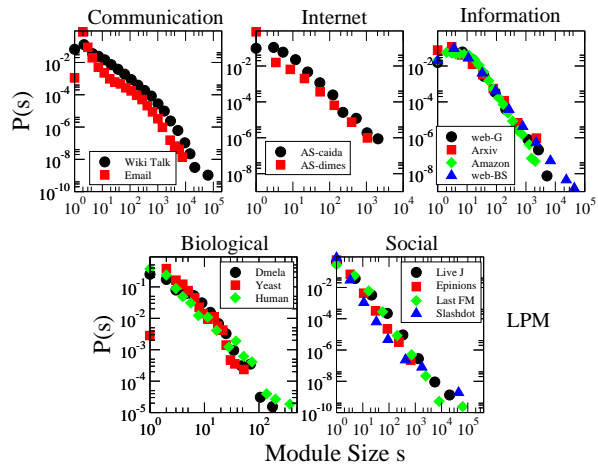}
\caption{Distribution of community sizes.}
\label{labelcomsize}
\end{center}
\end{figure}

\begin{figure}[h!]
\begin{center}
\includegraphics[width=\columnwidth]{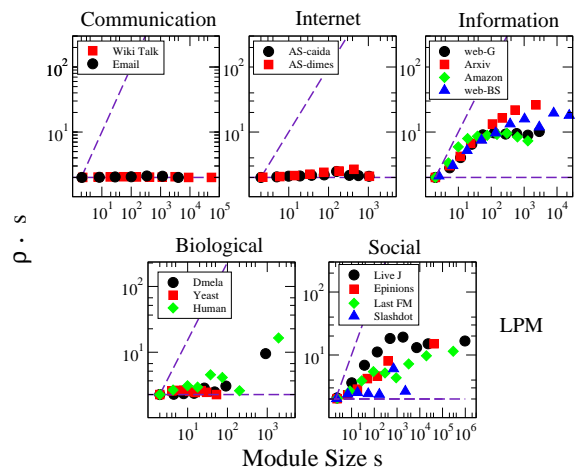}
\caption{Scaled link density of communities as a function of the community size.}
\label{labeldensity}
\end{center}
\end{figure}

\begin{figure}[h!]
\begin{center}
\includegraphics[width=\columnwidth]{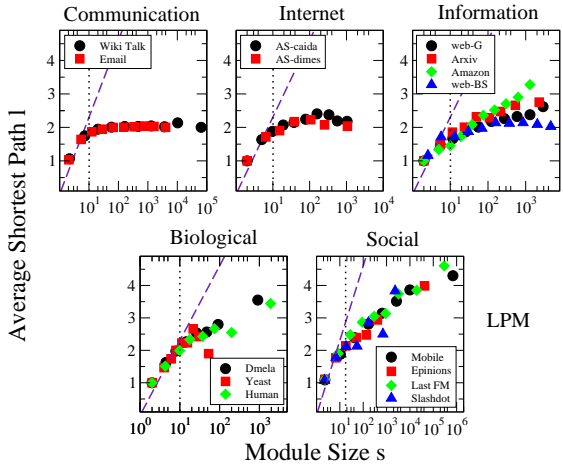}
\caption{ Average shortest path of a community as a function of the community size $s$.}
\label{labelsize_vs_asp}
\end{center}
\end{figure}

\begin{figure}[hb!]
\begin{center}
\includegraphics[width=\columnwidth]{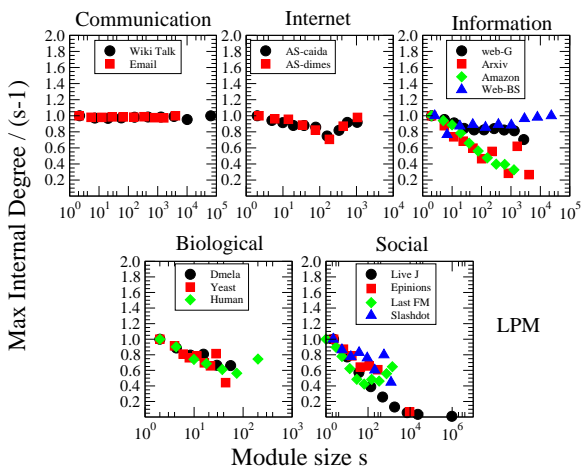}
\caption{Ratio between the maximum internal degree $\max(k_{in})$ of a
 node and the maximum possible number of internal neighbors $s-1$ as a function of $s$, the module size.}
\label{maxk}
\end{center}
\end{figure}

\begin{figure}[h!]
\begin{center}
\includegraphics[width=\columnwidth]{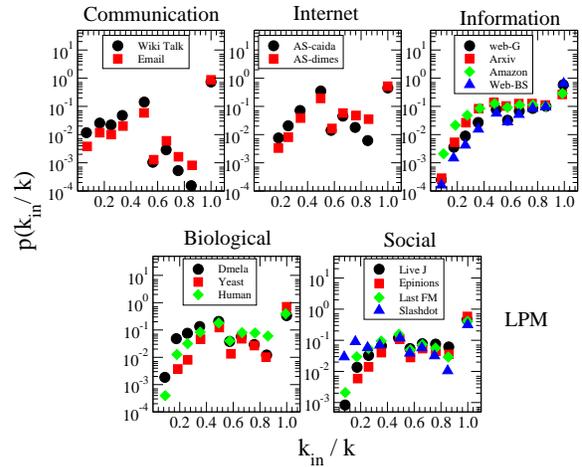}
\caption{ Distribution of the fraction of neighbors of a node belonging to the community of the node.}
\label{label_mu_distr_0}
\end{center}
\end{figure}

\section{Further Statistics on Community Properties}

In this section, we want to show some other statistical properties of
the modules. All figures display the results obtained 
using Infomap (upper panel) and the Label Propagation Method (lower panel).

\begin{figure}[h!]
\begin{center}
\includegraphics[width=\columnwidth]{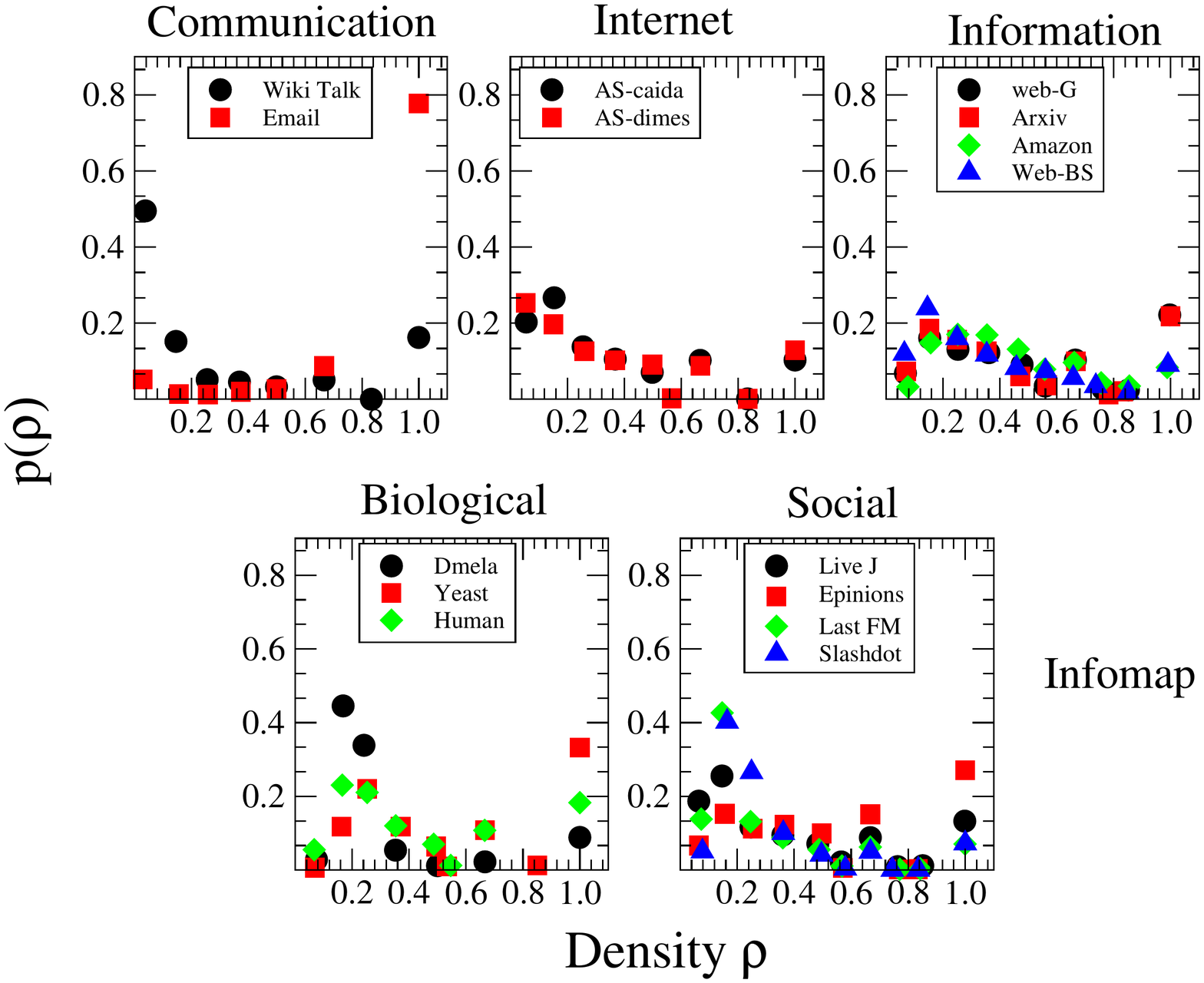}
\includegraphics[width=\columnwidth]{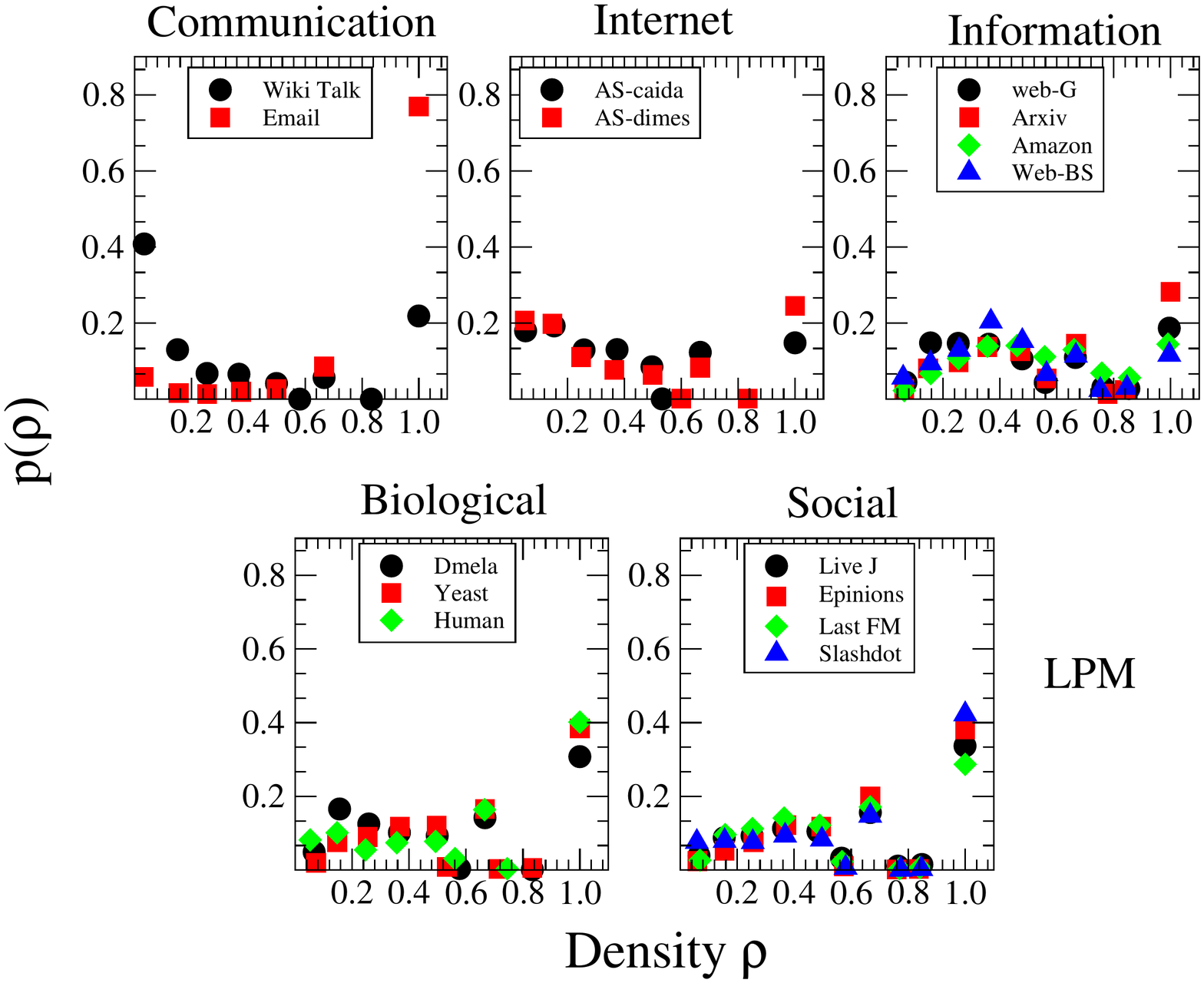}
\caption{Distribution of the link density for $s > 1$.}
\label{density_1}
\end{center}
\end{figure}

\begin{figure}[h!]
\begin{center}
\includegraphics[width=\columnwidth]{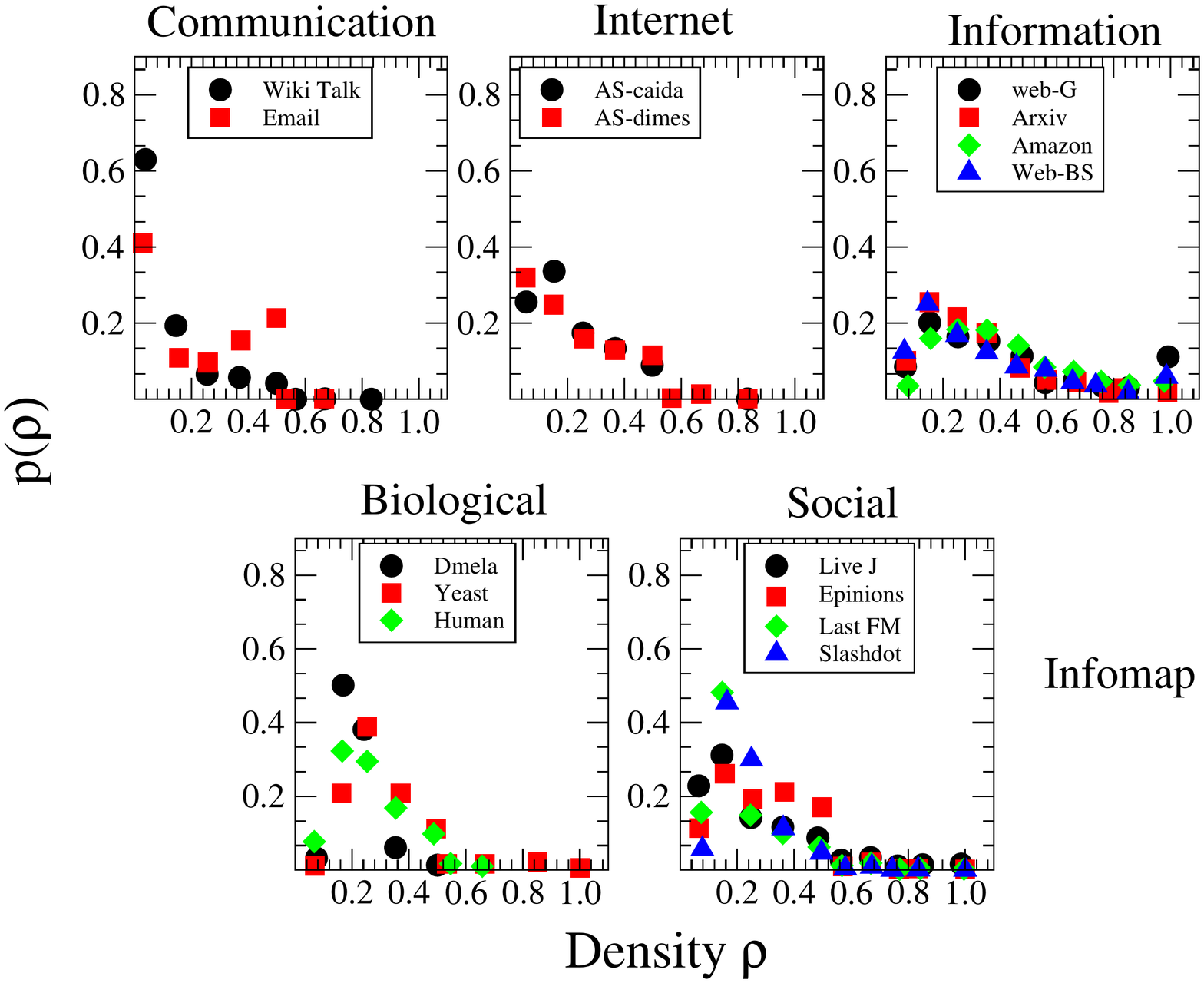}
\includegraphics[width=\columnwidth]{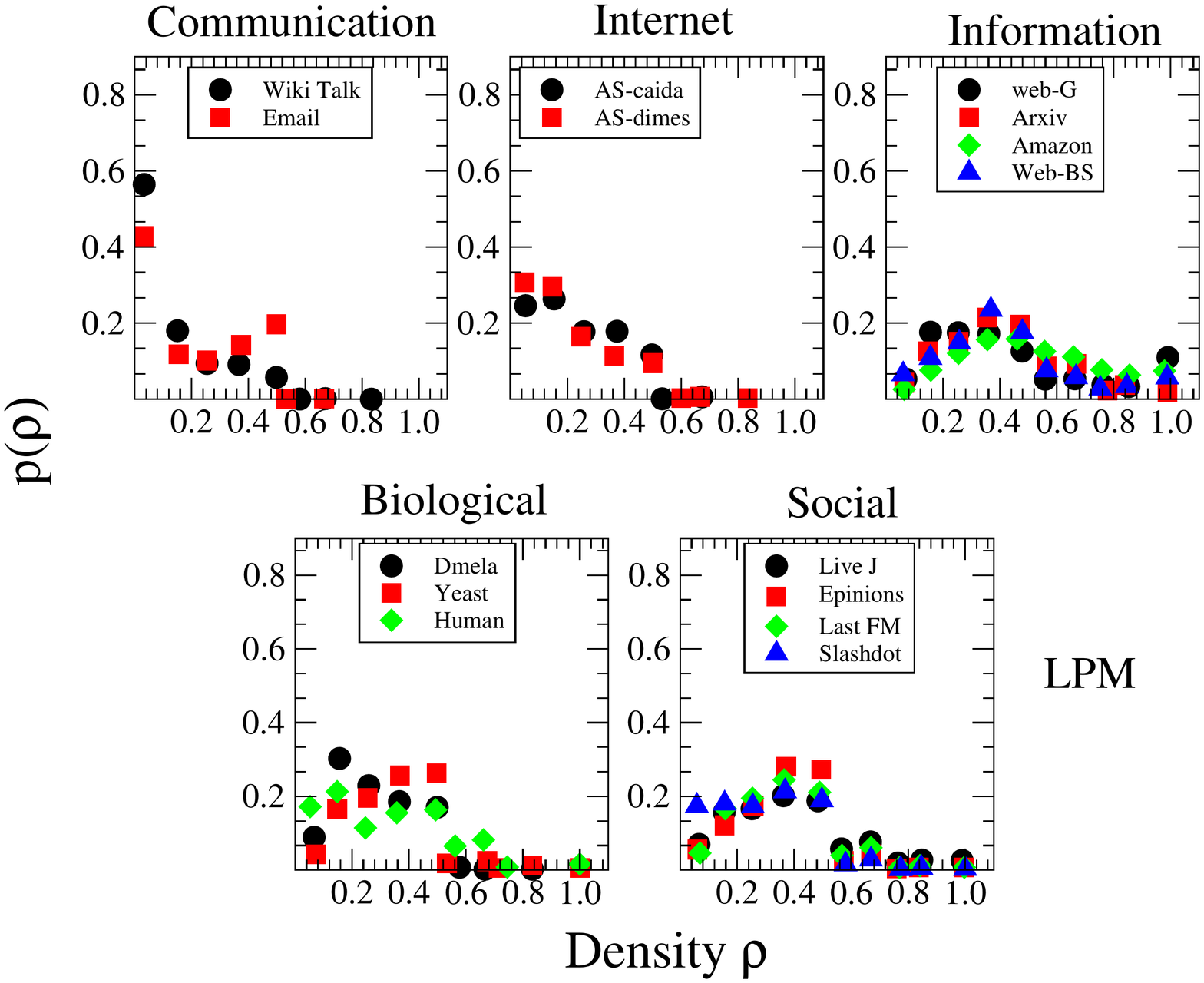}
\caption{Distribution of the link density  for $s > 3$.}
\label{density_3}
\end{center}
\end{figure}

\begin{figure}[h!]
\begin{center}
\includegraphics[width=\columnwidth]{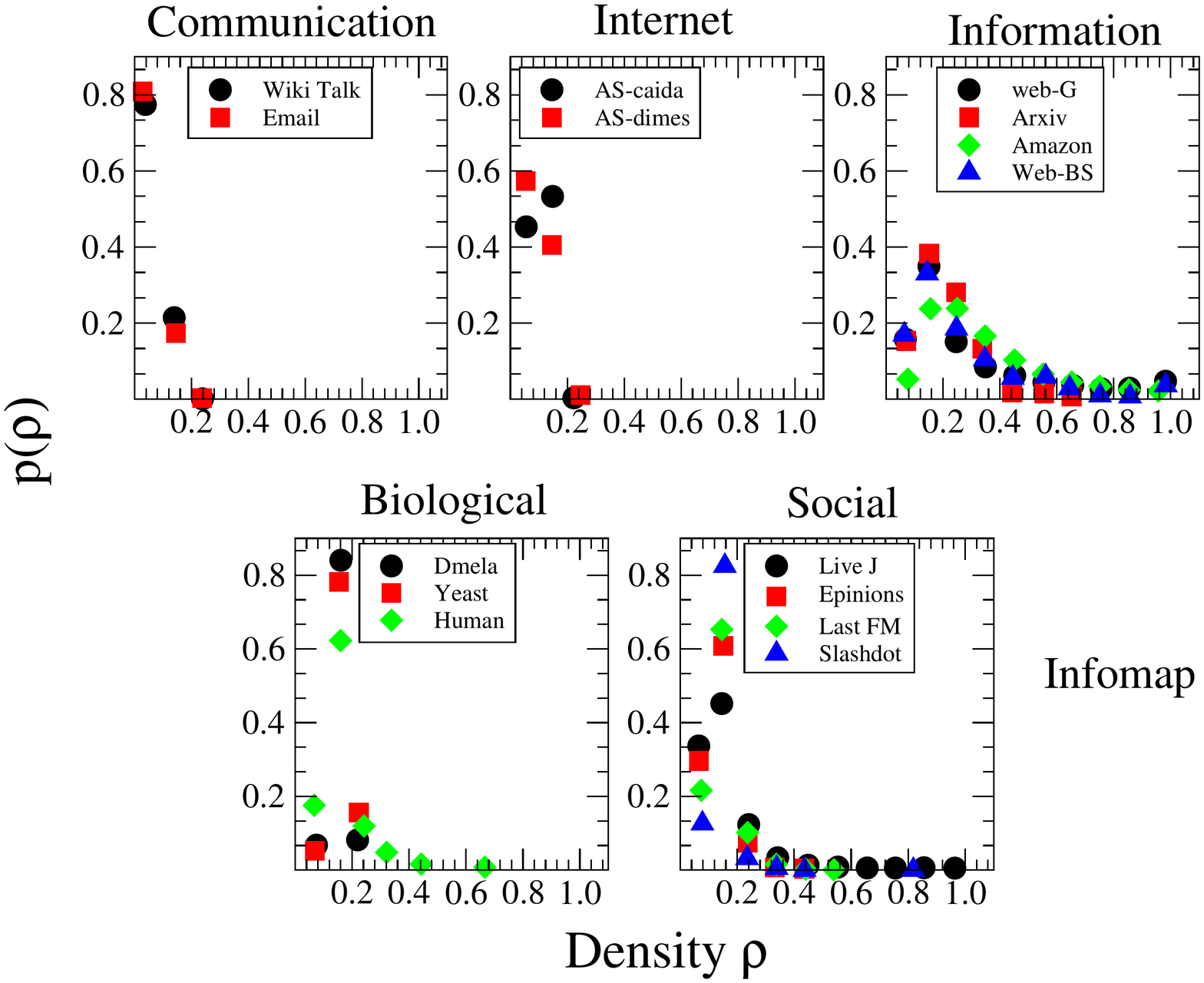}
\includegraphics[width=\columnwidth]{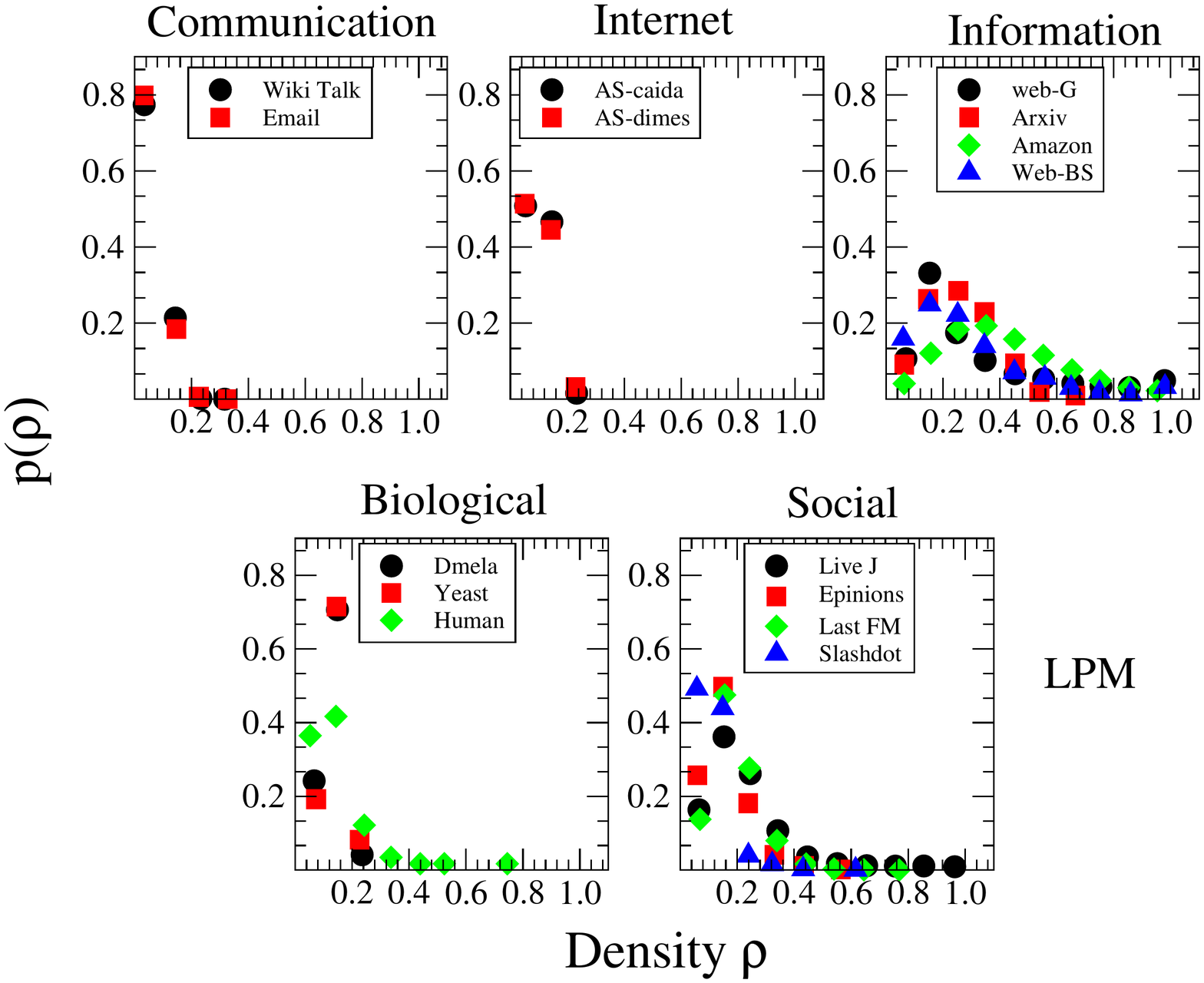}
\caption{Distribution of the link density  for $s > 10$.}
\label{density_10}
\end{center}
\end{figure}

As in the main article only the average values of link densities are shown, 
we first want to show what the probability distribution of the link
density $\rho$ of communities looks like. Fig. \ref{density_1} shows that in
all the systems there are dense modules together with sparser
modules. 
Nevertheless, there is a dependency on the size of the modules:
Fig. \ref{density_3} shows what happens if we 
discard very small communities, with less than $3$ nodes ($s < 3$), and
Fig. \ref{density_10} displays what is 
left when we consider fairly big modules, $s>10$; only social and
information networks include 
dense modules even after this filtering.

\begin{figure}[ht!]
\begin{center}
\includegraphics[width=\columnwidth]{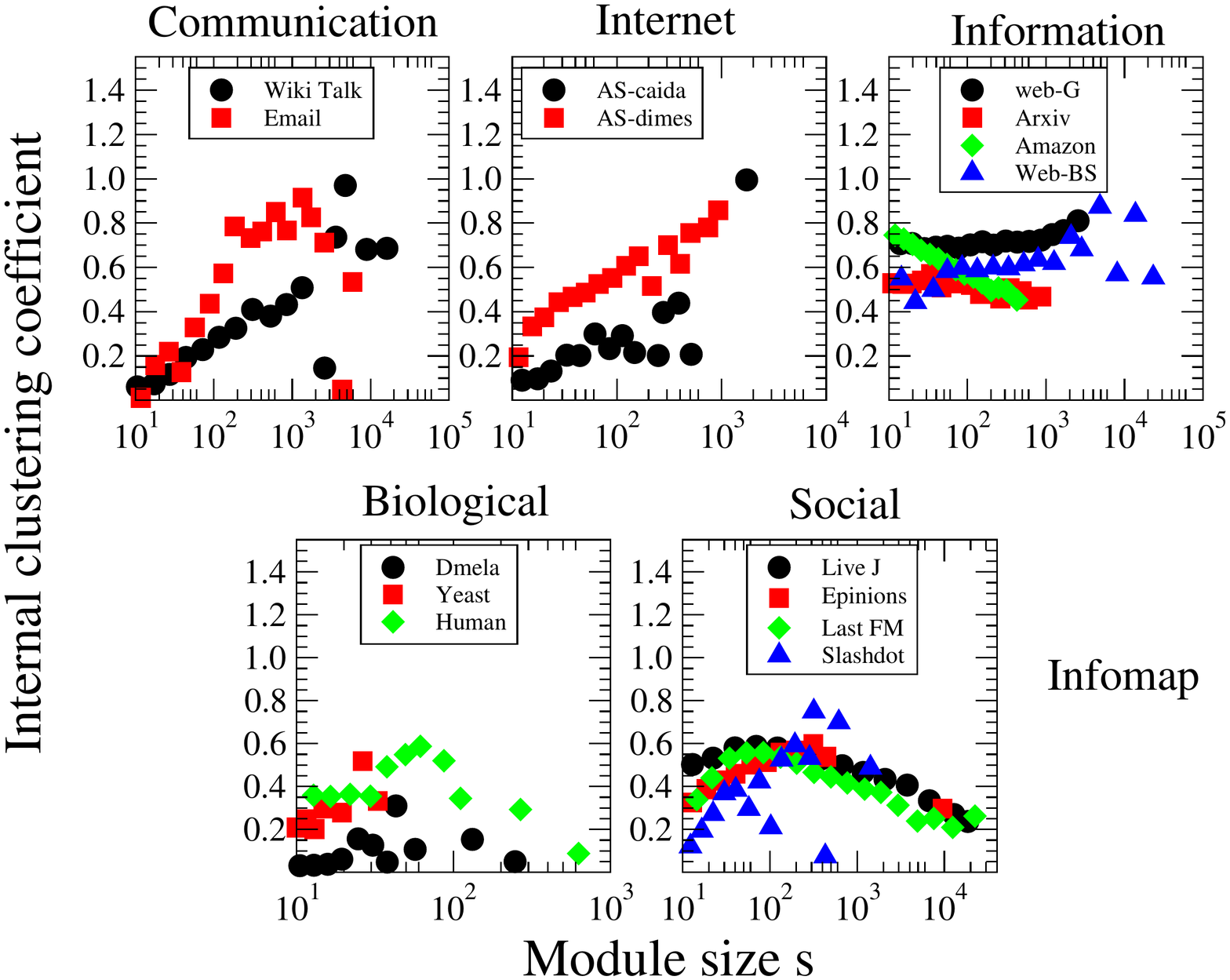}
\includegraphics[width=\columnwidth]{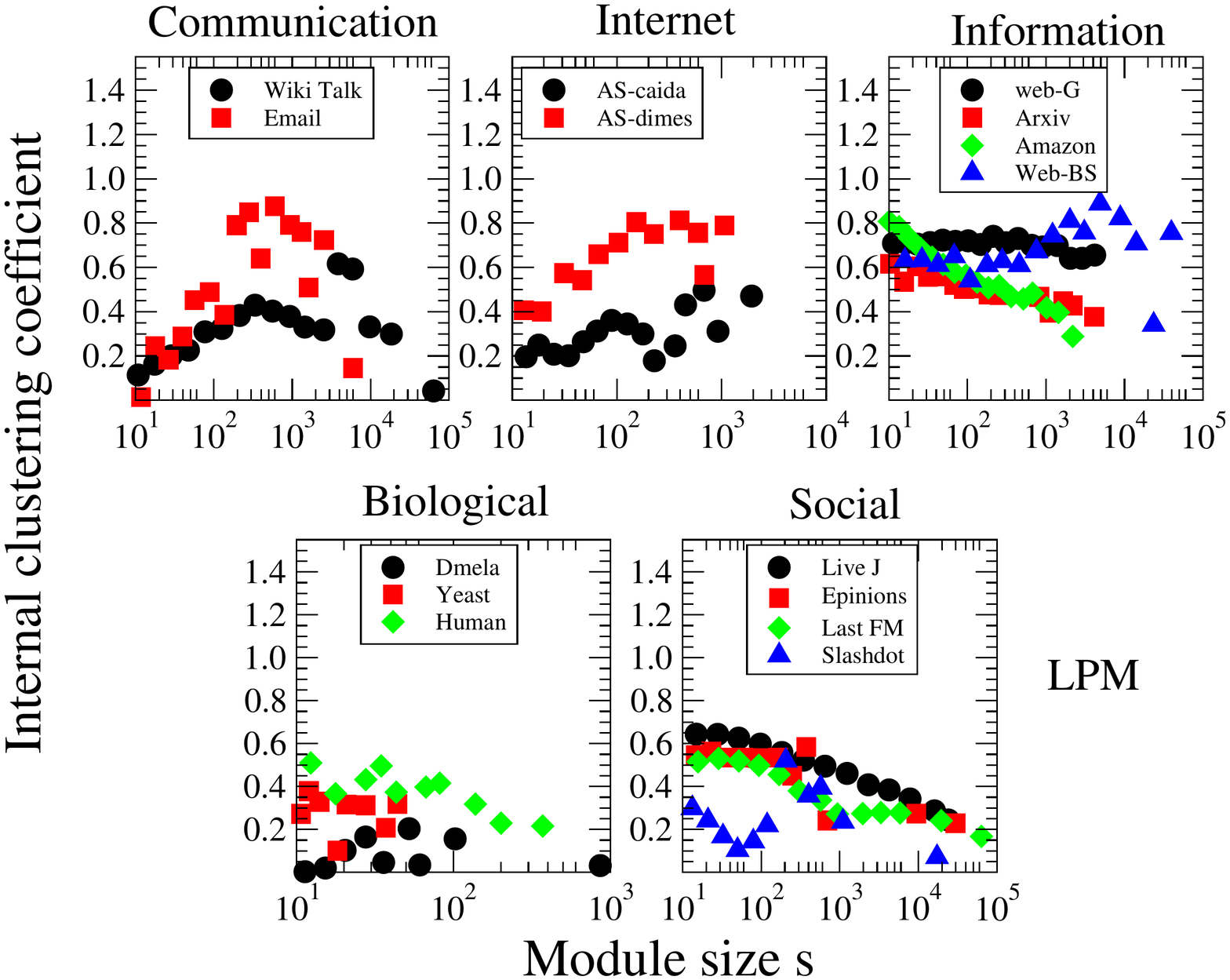}
\caption{Internal clustering coefficient as a function of the module size.}
\label{int_clu}
\end{center}
\end{figure}

Next, we show the average internal clustering coefficient as a function of
the module size $s$, Fig. \ref{int_clu}. The clustering coefficient $c$ is a node
property defined as the number of links between neighbors $t$ of the node divided
by the maximum possible number of such links for a node with the same degree $k$:
$c=t/(\frac{1}{2}k\left( k-1\right)) $. For nodes with degree smaller than two
we consider the clustering coefficient to be undefined and leave them out of the 
calculations of the averages.
Here, "internal" means that the
clustering coefficient is computed by only considering the subgraph of
the community, which includes only the internal links in the community. For communication systems and
the Internet, the average internal clustering coefficient of large communities can reach fairly high values 
although the corresponding densities $\rho$ are low. This can be explained in terms 
of "merged-star" structures, where two (or more) high-degree nodes are connected, their neighbours
have a low degree (approx.~the number of hubs) and are connected to all hubs. As then the clustering
coefficient for these nodes is typically unity and their number is large, they dominate the average 
clustering coefficient within the community.

\section{Further Statistics of Node Properties}

Here we focus on properties of nodes with respect to their
communities. Again, we show results obtained using both Infomap (upper panel) and
the Label Propagation Method (lower panel).

\begin{figure}[h!]
\begin{center}
\includegraphics[width=\columnwidth]{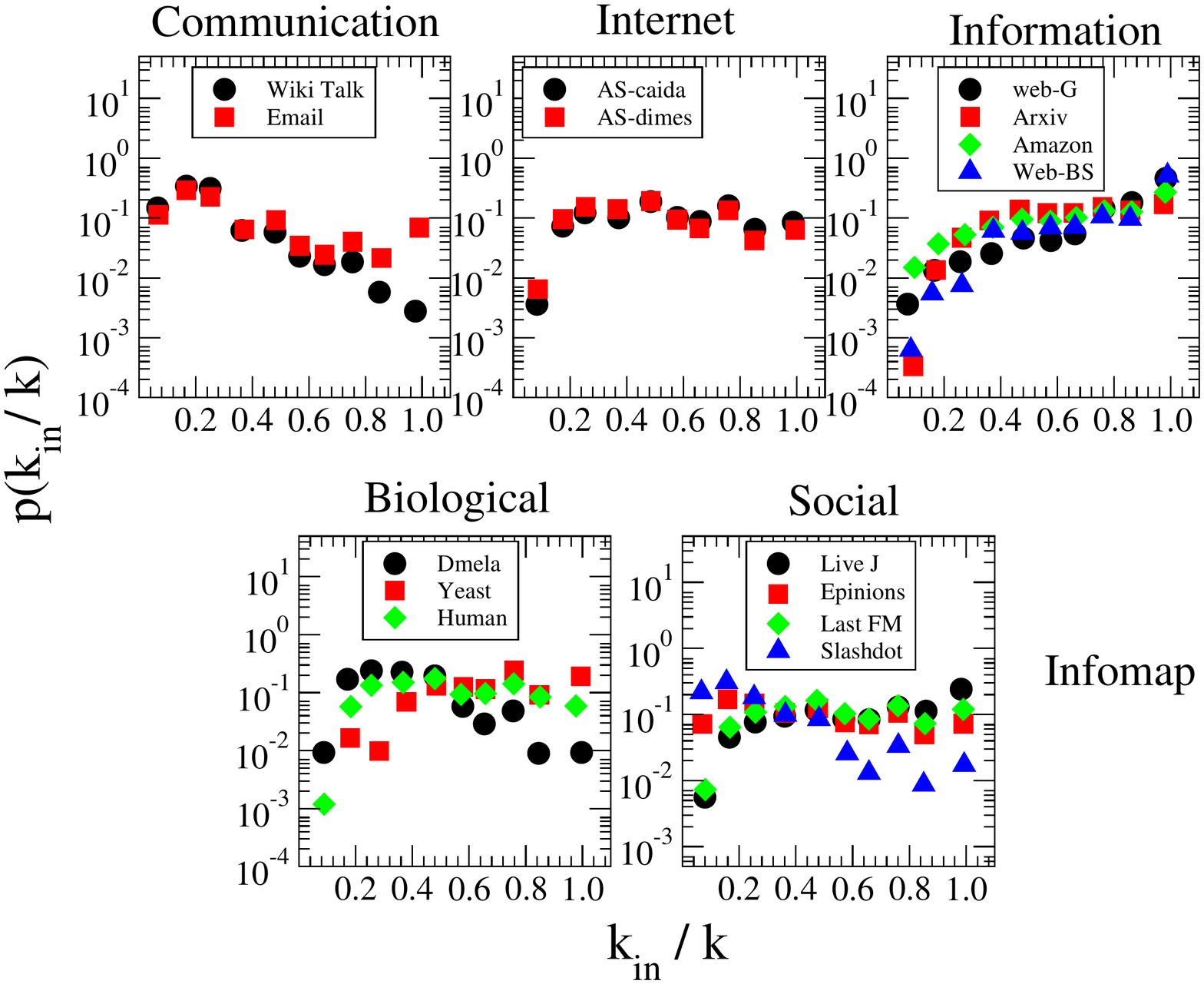}
\includegraphics[width=\columnwidth]{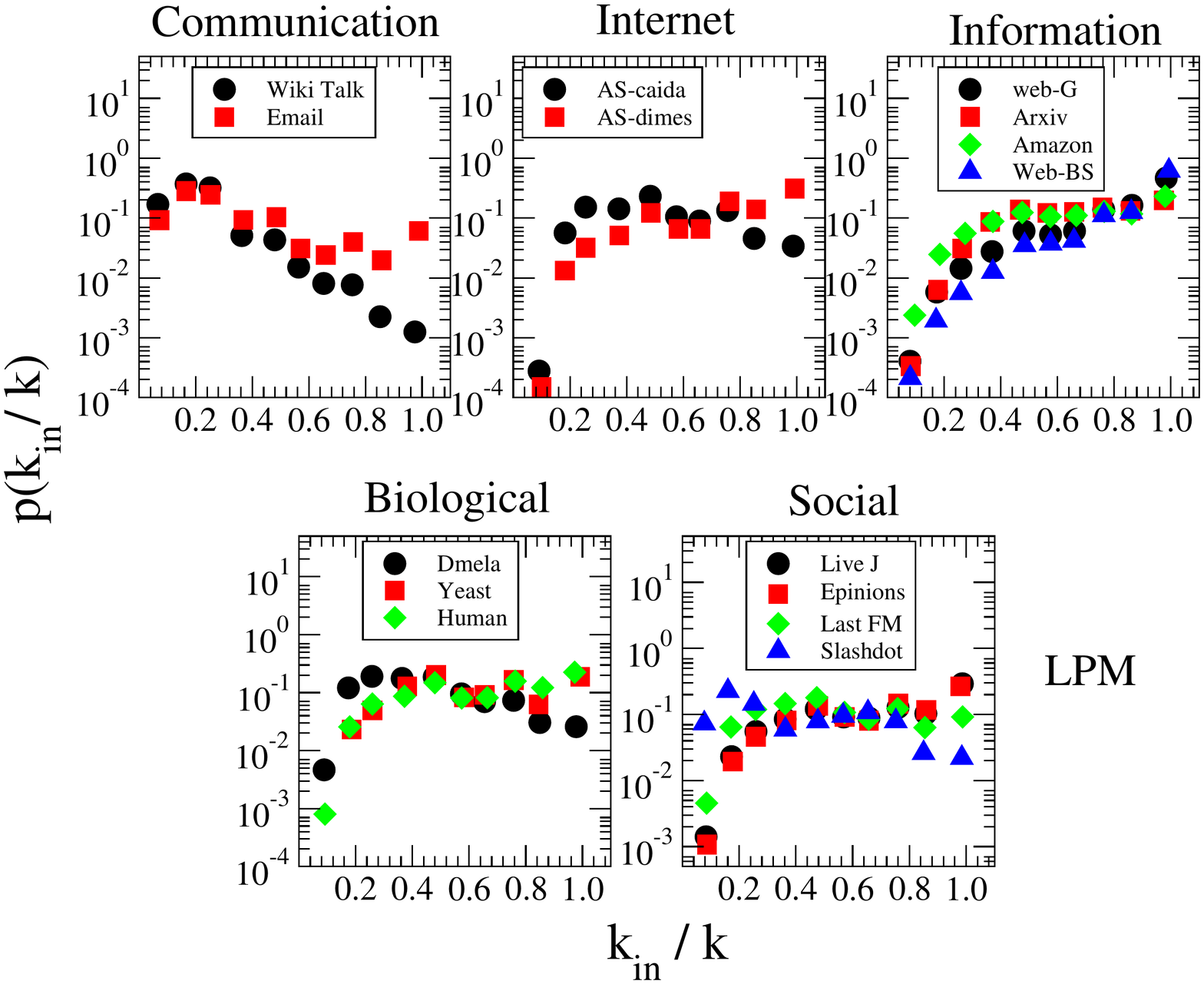}
\caption{Distribution of $k_{in}/k$, for $k>3$.}
\label{mu_distr_3}
\end{center}
\end{figure}

\begin{figure}[h!]
\begin{center}
\includegraphics[width=\columnwidth]{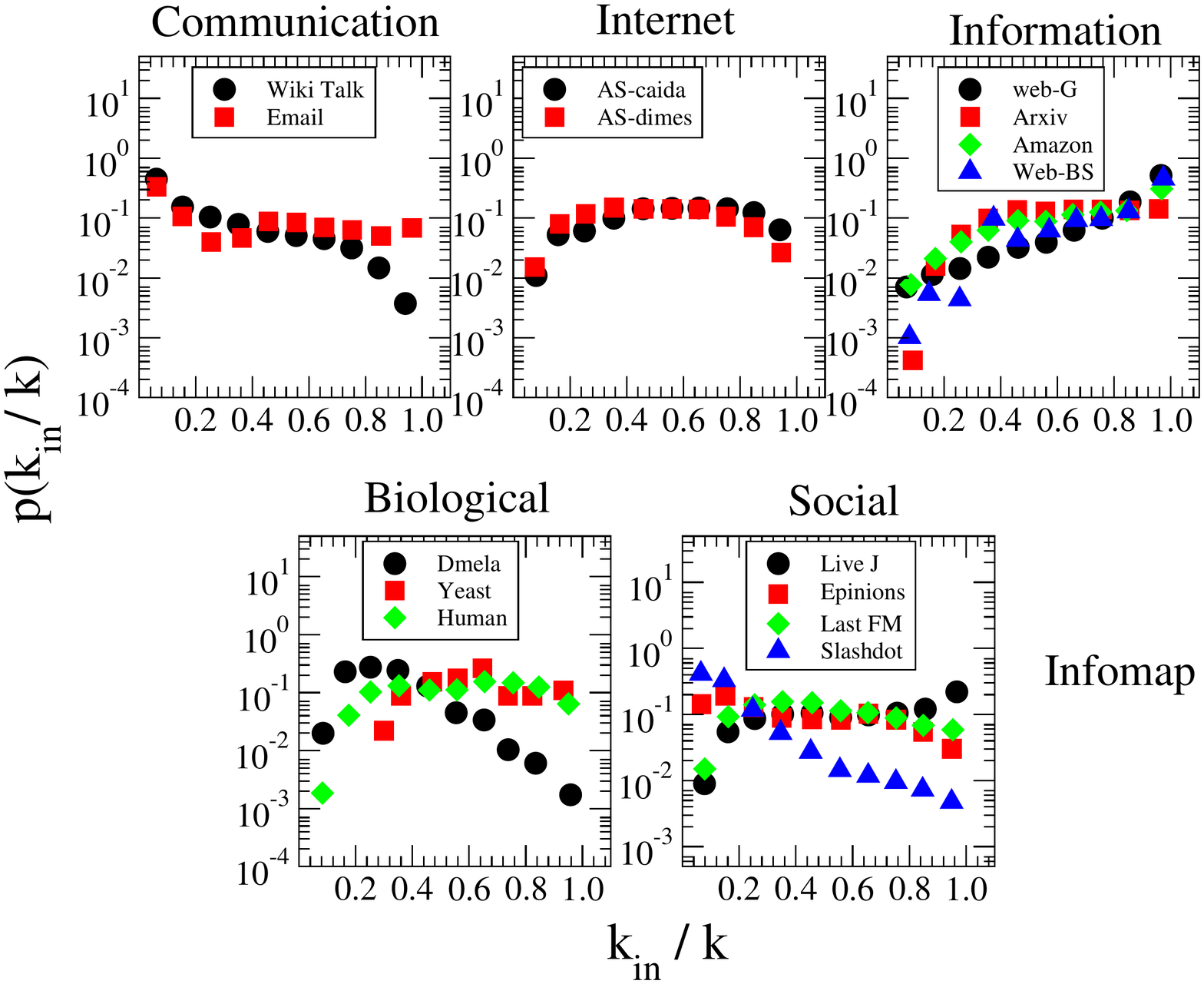}
\includegraphics[width=\columnwidth]{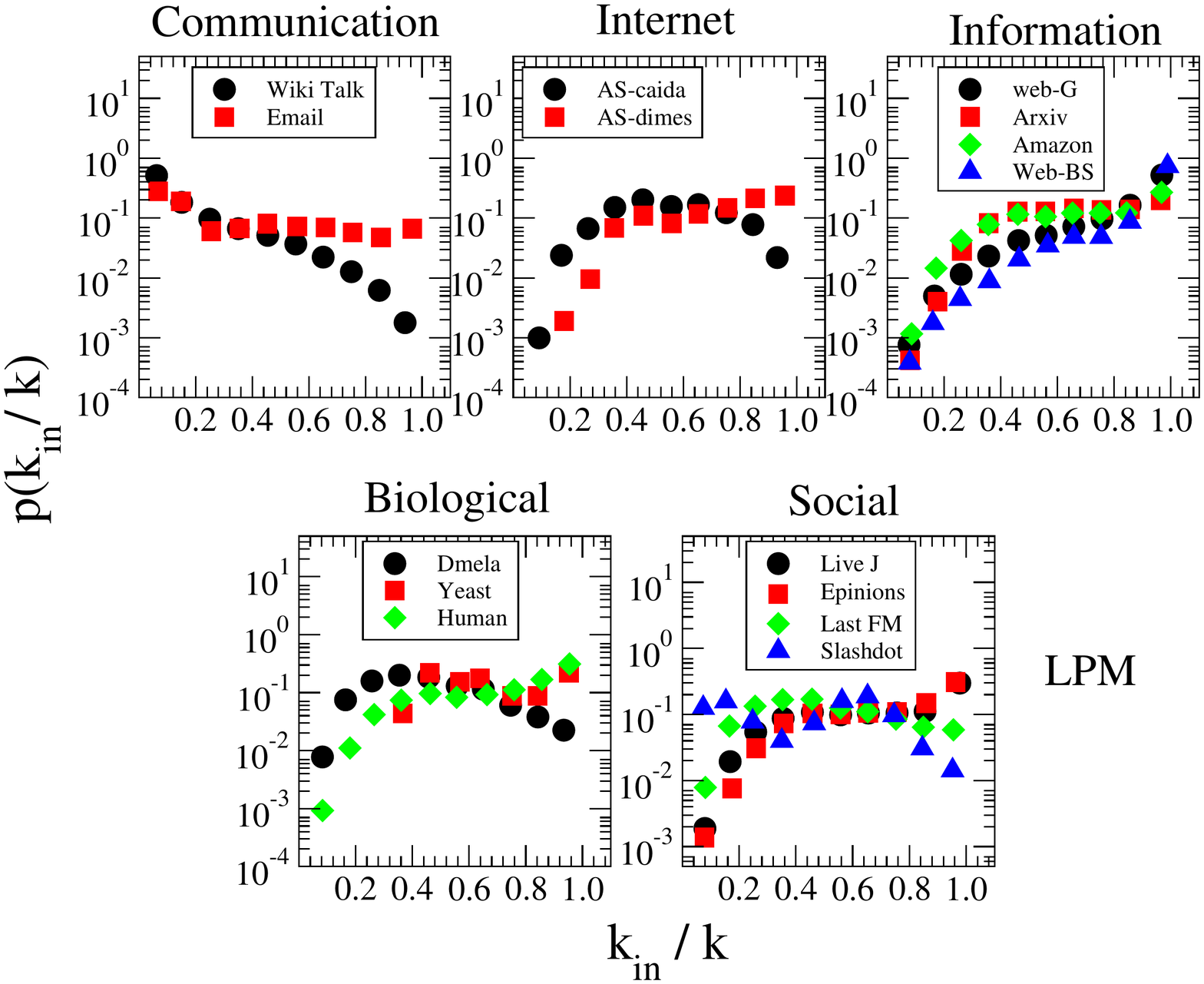}
\caption{Distribution of $k_{in}/k$, for $k>10$.}
\label{mu_distr_10}
\end{center}
\end{figure}

In Fig. \ref{mu_distr_3}, we show the distribution of the fraction of
neighbors of a node belonging to its community  
when one only considers nodes with degree $k > 3$ (in the main
manuscript we considered all nodes). Fig. \ref{mu_distr_10} is the
same plot, but including only nodes with degrees larger than $10$. The
two plots display flatter curves than the full distributions, and they
look much smoother, indicating that the fluctuations observed in the full
curves are mostly due to low degree nodes. Low degree nodes can cause peaks
in the plots because the values of the fraction $k_{in}/k$ are quantized
(\emph{e.g.} for a node of degree two the fraction must be $0$, $0.5$ or $1$).
By observing the
rightmost points of the curves, we see that they lie much lower than
the corresponding points of the full curves, except for the
information networks. This means that many
nodes which are fully embedded in their community have low degree, as
expected.

\clearpage

\begin{table*}[htb!]
\center
\begin{tabular}{ | l | l | c | c  |  c | c | }
\hline
\multicolumn{6}{|c|}{Network degree distribution} \\ \hline
\multirow{1}{*}{Category}
& name & exponent  & min degree & exp error & $p-$value \\ 
\hline
\multirow{2}{*}{Communication}
& wikitalk & -2.46 & 1 & 0.01 & 0 \\
& email & -2.93 & 1 & 0.01 & 0 \\
 \hline

\multirow{2}{*}{Internet} 
&caida & -2.12 & 5 & 0.03 & 0.6 \\
&dimes & -2.2 & 2  & 0.01 & 0.2 \\
\hline

\multirow{4}{*}{Information} 
&Web Google & -2.68 & 23 & 0.01 & 0.8 \\
&arxiv & -3.19 & 61 & 0.04 & 0.3 \\
&amazon & -3.27 & 17 & 0.03 & 0\\
&Web BS & -2.59 & 46 & 0.02 & 0 \\ 
\hline

\multirow{3}{*}{Biological} 
&dmela & - 3.50 & 28 & 0.05 & 0 \\
&yeast & -3.0 & 5 & 0.5 & 0.4  \\
&human & 3.0 & 31 & 0.2 & 0.1 \\
\hline

\multirow{4}{*}{Social} 
&live j. & -2.8 & 86 & 0.1 & 0.3 \\
&epinions & -1.70 & 1 & 0.01 & 0 \\
&last fm & -2.9 & 35  & 0.1 & 0\\
&slashdot & -2.5 & 43 & 0.5 & 0.8  \\
\hline
\end{tabular}
\caption{\label{table10} Power-law exponents of the degree distribution and the
minimum degree from which the fit holds. 
We used maximum likelihood fitting \cite{clauset_maxl}.}
\end{table*}

\begin{table*}[htp!]
\center
\begin{tabular}{ | l | l | c | c  | c | c |  }
\hline
\multicolumn{6}{|c|}{Community size distribution (from Infomap)} \\ \hline
\multirow{1}{*}{Category}
& name & exponent  & min degree & exp error & $p-$value \\ 
\hline
\multirow{2}{*}{Communication}
& wikitalk & -2.7 & 881 & 0.3 & 0.1\\
& email & -2.8 & 674 & 0.3 & 0.5\\
 \hline

\multirow{2}{*}{Internet} 
&caida & -2.10 & 11 & 0.05 & 0.4\\
&dimes & -2.00 & 18 & 0.05 & 0.9 \\
\hline

\multirow{4}{*}{Information} 
&Web Google & -2.57 & 89 & 0.03 & 0.2\\
&arxiv & -2.4 & 69 & 0.3 & 0.5 \\
&amazon & -3.5 & 97 & 0.2 & 0.02\\
&Web BS & -2.4 & 36 & 0.1 & 0\\ 
\hline

\multirow{3}{*}{Biological} 
&dmela & - 3.5 & 9 & 0.1 & 0.1\\
&yeast & -3.05 & 8 & 0.05 & 0.1\\
&human & 2.6 & 8 & 0.1 & 0.1\\
\hline

\multirow{4}{*}{Social} 
&live j. & -2.22 & 59 & 0.02 & 0\\
&epinions & -2.5 & 13 & 0.2 & 0.3\\
&last fm & -2.70 & 34 & 0.05 & 0 \\
&slashdot & -3.5 & 10 & 0.1 & 0 \\
\hline
\end{tabular}
\caption{\label{table20} Power-law exponents of the community size distribution
derived from Infomap. 
We used maximum likelihood fitting \cite{clauset_maxl}.}
\end{table*}

\begin{table*}
\center
\begin{tabular}{ | l | l | c | c  | c | c |  }
\hline
\multicolumn{6}{|c|} {Community size distribution (from LPM)} \\ \hline
\multirow{1}{*}{Category}
& name & exponent  & min degree & exp error & $p-$value \\ 
\hline
\multirow{2}{*}{Communication}
& wikitalk & -2.6 & 1145 & 0.2 & 0.4\\
& email & -2.4 & 248 & 0.1 & 0.2 \\
 \hline

\multirow{2}{*}{Internet} 
&caida & -2.08 & 13 & 0.08 & 0.4\\
&dimes & -1.95 & 12 & 0.05 & 0.8\\
\hline

\multirow{4}{*}{Information} 
&Web Google & -2.45 & 36 & 0.02 & 0.1 \\
&arxiv & -2.0 & 16  & 0.1 & 0.1\\
&amazon & -2.80 & 30 & 0.05 & 0.3\\
&Web BS & -2.0 & 107 & 0.1 & 0.7 \\ 
\hline

\multirow{3}{*}{Biological} 
&dmela & - 2.7 & 10 & 0.5 & 0.3\\
&yeast & -2.6 & 5 & 0.5 & 0.3\\
&human & 1.9 & 2 & 0.05 & 0.2 \\
\hline

\multirow{4}{*}{Social} 
&live j. & -2.40 & 86  & 0.05 & 0.1\\
&epinions & -2.40 & 5 & 0.05 & 0.2 \\
&last fm & -2.9 & 35 & 0.1 & 0.1 \\
&slashdot & -2.7 & 24 & 0.1 & 0 \\
\hline
\end{tabular}
\caption{\label{table30} Power-law exponents of the community size distribution
derived from the LPM. 
We used maximum likelihood fitting \cite{clauset_maxl}.}
\end{table*}
%
\end{appendix}

\end{document}